\documentclass[conference]{IEEEtran}

\usepackage[noadjust]{cite} 

\usepackage{graphicx}
\graphicspath{ {./figs/} }
\usepackage{subcaption} 

\usepackage{amsmath}
\usepackage{amsfonts}  

\usepackage{algorithm}
\usepackage{algorithmic}

\usepackage[table]{xcolor} 
\usepackage{tabularx}
\usepackage{multirow} 
\usepackage{adjustbox} 
\usepackage{wrapfig}

\usepackage[misc]{ifsym} 

\usepackage[colorlinks=true, linkcolor=magenta, urlcolor=blue, citecolor=cyan, bookmarks=true, bookmarksopen, pagebackref=true]{hyperref} 
\title{How to Backdoor Consistency Models?}

\author{\IEEEauthorblockN{Chengen Wang$^{(\textrm{\Letter})}$}
\IEEEauthorblockA{University of Texas as Dallas\\
Email: chengen.wang@utdallas.edu}
\and
\IEEEauthorblockN{Murat Kantarcioglu}
\IEEEauthorblockA{University of Texas as Dallas\\
    Email: muratk@utdallas.edu}}

\begin{document}
\maketitle

\begin{abstract}
Consistency models are a new class of models that generate images by directly mapping noise to data, allowing for one-step generation and significantly accelerating the sampling process. However, their robustness against adversarial attacks has not yet been thoroughly investigated. In this work, we conduct the first study on the vulnerability of consistency models to backdoor attacks. While previous research has explored backdoor attacks on diffusion models, those studies have primarily focused on conventional diffusion models, employing a customized backdoor training process and objective, whereas consistency models have distinct training processes and objectives. Our proposed framework demonstrates the vulnerability of consistency models to backdoor attacks. During image generation, poisoned consistency models produce images with a Fréchet Inception Distance (FID) comparable to that of a clean model when sampling from Gaussian noise. However, once the trigger is activated, they generate backdoor target images. We explore various trigger and target configurations to evaluate the vulnerability of consistency models, including the use of random noise as a trigger. This novel trigger is visually inconspicuous, more challenging to detect, and aligns well with the sampling process of consistency models. Across all configurations, our framework successfully compromises the consistency models while maintaining high utility and specificity. We also examine the stealthiness of our proposed attack, which is attributed to the unique properties of consistency models and the elusive nature of the Gaussian noise trigger. Our code is available at \href{https://github.com/chengenw/backdoorCM}{https://github.com/chengenw/backdoorCM}.

\end{abstract}


\section{Introduction} \label{sec:intro}

\begin{figure}[h] 
     \centering
     \includegraphics[width=0.45\textwidth]{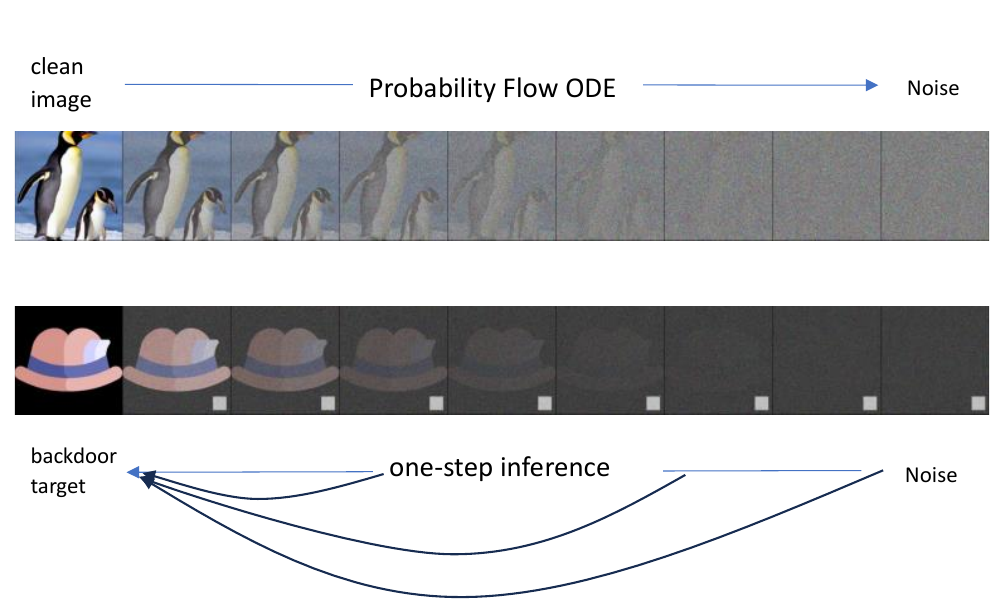} 
     \caption{Training backdoor consistency model using both clean images and backdoor targets. In this example, the target is a hat and the trigger is a square box at the bottom right corner.}
     \label{fig:noise-schedule}
\end{figure}

In recent years, diffusion models \cite{sohl2015deep_dm,song2019generativeModeling,DDPM,song2021scorebased} have significantly advanced deep learning-based generation techniques across multiple domains, including image, audio \cite{kong2020diffwave_dm_audio} and video \cite{ho2022video_dm}. One disadvantage of these models is that they generate samples through hundreds or thousands of iterative steps, which require much longer processing time compared to previous generative models such as GAN \cite{GAN} or VAE \cite{vae}. Although researchers have made efforts to address this issue, diffusion models still require numerous steps to generate high-quality images using improved sampling techniques \cite{liu2022pseudo,lu2022dpm_solver}, or require retraining through the distillation of pre-trained diffusion models~\cite{salimans2022progressive_distill}. 

To address the above challenge, \cite{song2023consistency} proposes consistency models (CM), which can generate high-quality samples in one step. They also support multi-step sampling, allowing for a trade-off between computation and quality. Although consistency models can be trained by the distillation of diffusion models, they can also be trained independently. This make consistency models a new family of generative models.

Given the advantages of consistency models and their potential for a wide range of applications, we seek to gain a deeper understanding of their security implications, as these models can serve as foundational building blocks for developing downstream applications~\cite{rombach2022highresolutionimagesynthesislatent}. In this work, we focus on the vulnerability of consistency models to backdoor attack within the image domain. If these base models are compromised with a backdoor during training, they can generate unintended or malicious images when the backdoor trigger is activated, potentially leading to serious consequences.

Our objective is to explore how consistency models can be backdoored and to evaluate their performance across various backdoor settings. While several studies have examined the vulnerabilities of diffusion models to backdoor attacks \cite{chen2023trojdifftrojanattacksdiffusion,chou2023backdoordiffusionmodels,chou2024villandiffusion}, they primarily concentrate on basic diffusion models and do not cover consistency models, which have distinct training processes and objectives. As a result, the potential risks associated with consistency models remain under-explored.

Backdoor attacks on clean models typically have two primary goals. First, the tampered model should function normally and produce clean images when the backdoor is not triggered, ensuring high utility. Second, the tampered model should generate target images, as defined by the attacker, once the hidden trigger is activated, ensuring high specificity.

Figure \ref{fig:noise-schedule} illustrates how the backdoor consistency model is trained, while Figure \ref{fig:progress-noise-hat-DM} shows how sampling on the compromised model behaves before and after the trigger is activated.

In previous work on diffusion model backdoor attacks, common triggers include special shapes or objects,  such as a box or glasses. In this work, we propose also using noise as a trigger, leveraging the fact that consistency models generate images by sampling from Gaussian noise, making the noise trigger less noticeable. The noise trigger is a sample from a Gaussian distribution, which remains fixed during both the model backdoor training and inference stages.

Our main contributions are as follows:
\begin{enumerate}
    \item To the best of our knowledge, we are the first to propose a backdoor attack tailored to consistency models, 
    \item We conducted extensive experiments to reveal the vulnerabilities of consistency models under various trigger, target, and poison rate settings.
    \item We propose using Gaussian noise as a trigger, which is less noticeable and well-suited to the sampling process of consistency models. This trigger is easy to generate and difficult to detect.
\end{enumerate}

Our paper is organized as follows: we begin by introducing the background and motivation for investigating the vulnerability of consistency models in Section \ref{sec:intro}, followed by a discussion of related work in Section \ref{sec:related}. Next, we present the preliminary knowledge necessary for the understanding of this work in Section \ref{sec:preliminary}, and describe how to backdoor consistency models in Section \ref{sec:methods}. We detail the experiments in Section \ref{sec:exp}, and discuss the stealthiness of the proposed attack in Section \ref{defense}. Finally, we address the limitations of our work in Section \ref{sec:limit} and conclude the paper in Section \ref{sec:conclusion}.

\section{Related Work} \label{sec:related}

\subsection{Consistency Models}
Consistency models \cite{song2023consistency,song2023improved_cm} significantly accelerate the sampling process of diffusion models, enabling one-step sample generation. These models have been extended in various ways. For instance, \cite{kim2023consistency_traject,heek2024multistep_cm} extend the models to multi-step sampling, while \cite{xiao2023ccm_conditional} introduces conditional consistency models. Additionally \cite{luo2023latent_cm} extends the approach to latent space. Notably, \cite{geng2024consistencymodelseasy} substantially improves the training efficiency of consistency models by starting from a pre-trained diffusion model and progressively refining the full consistency conditions throughout the training process.

\subsection{Backdoor Attacks on Diffusion Models}
\cite{chou2023backdoordiffusionmodels,chen2023trojdifftrojanattacksdiffusion} are the first work to investigate backdoor attack on diffusion models. They introduced a customized training schedule specifically designed to backdoor denoising diffusion models \cite{DDPM}. \cite{chou2024villandiffusion} extends backdoor attack to a more generalized version, although it is still not a universal one. Their work focuses on standard diffusion models, analyzing the reversed process primarily in the form of $q(x_{t-1}|x_t, x_0)$, similar to the approach in \cite{DDPM}, although with a more generalized training schedule. Additionally, it does not provide an explicit correction term for a universal backdoor training schedule.

To make the trigger less noticeable, \cite{li2024invisible_backdoor} proposes using random noise as a trigger through a bi-level optimization approach. While this trigger is similar to the noise trigger introduced in our work, our approach involves directly sampling from a Gaussian distribution and fixing this noise trigger during backdoor training. This choice is based on our understanding that any Gaussian noise sample is generally inconspicuous in consistency model inference settings.

\section{Preliminaries} \label{sec:preliminary}
\subsection{Diffusion Models}
Since consistency models are inspired by the theory of diffusion models, we first introduce diffusion models. Let us denote the data distribution by $p_{data}(\mathbf{x})$. The diffusion process perturbs data via Gaussian perturbations and is described by a stochastic differential equation (SDE) \cite{song2021scorebased}
\begin{equation}
    \mathrm{d}\mathbf{x}_t = \boldsymbol{f}(\mathbf{x},t)\mathrm{d}t + g(t)\mathrm{d}\mathbf{w},
\end{equation}
where $t\in[0,T]$, $T>0$ is a predefined constant, set to 80 in \cite{karras2022elucidating}, $\boldsymbol{f}(\cdot,t)$ is the drift coefficient, $g(\cdot)$ the diffusion coefficient of $\mathbf{x}_t$, and $\mathbf{w}$ represents the standard Brownian motion. This SDE has a corresponding ordinary differential equation (ODE) known as Probability Flow (PF) ODE \cite{song2021scorebased}, which shares the same marginal probability $p_t(\mathbf{x})$:
\begin{equation}
\mathrm{d}\mathbf{x}_t = \left[ \boldsymbol{f}(\mathbf{x},t)-\frac{1}{2}g(t)^2\nabla_\mathbf{x}\log p_t(\mathbf{x}) \right]\mathrm{d}t,
\end{equation}
where the term $\nabla_\mathbf{x}\log p_t(\mathbf{x})$ is known as the score function \cite{song2019generativeModeling,song2021scorebased}.

When adopting a Gaussian perturbation schedule defined as $\mathbf{x}_t\sim \mathcal{N}(\mathbf{x};t^2\boldsymbol{I})$ \cite{karras2022elucidating}, where $\mathbf{x}\sim p_{data}$, the PF ODE is simplified as
\begin{equation}
    \mathrm{d}\mathbf{x}_t=-t\nabla_\mathbf{x}\log p_t(\mathbf{x})\mathrm{d}t.
\end{equation}

\subsection{Consistency Models}
Consistency models are a new type of models designed for single-step generation. These models are trained to map any point on a PF ODE trajectory back to the origin of the same trajectory. For a given solution trajectory of the PF ODE, the consistency function is defined as $\boldsymbol{f}:(\mathbf{x}_t,t)\mapsto\mathbf{x}$, where $t\in[\epsilon,T]$. This means the outputs are consistent for any pairs of $(\mathbf{x}_t,t)$ on the same PF ODE trajectory \cite{song2023consistency}.

As shown in \cite{song2023consistency}, an unbiased estimator for the score function in the PF ODE is
\begin{equation}
    \nabla \log p_t(\mathbf{x}_t)=-\mathbb{E}\left[ \frac{\mathbf{x}_t - \mathbf{x}}{t^2}\vline\mathbf{x}_t\right],
\end{equation}
thereby eliminating the need to learn a score function during consistency model training. This allows consistency models to be trained independently without relying on the distillation of a pre-trained diffusion model.

The consistency training loss in \cite{song2023consistency} is defined as follows:
\begin{align}\label{eq:loss}
&   \mathcal{L}_{CT}^N(\boldsymbol{\theta},\boldsymbol{\theta^-}) =\notag \\
&    \mathbb{E}[\lambda(t_n)d(\boldsymbol{f_\theta}(\mathbf{x}_{n+1}, t_{n+1}),\boldsymbol{f_{\theta^-}}(\mathbf{x}_n,t_n))],
\end{align}
where $\lambda(\cdot)$ is a positive weighting function, $d(\cdot)$ is a distance function, $\boldsymbol{\theta}$ represents the model parameters, and $\boldsymbol{\theta^-}$ is the running average of the past values of $\boldsymbol{\theta}$ during training. The function $\boldsymbol{f_\theta}(\cdot,\cdot)$ is a learned consistency function, while $\mathbf{x}_{n+1}$ and $\mathbf{x}_n$ are two adjacent samples at times $t_{n+1}$ and $t_n$, respectively. These samples are defined as $\mathbf{x}_{n+1}=\mathbf{x}+t_{n+1}\boldsymbol{\epsilon}$ and $\mathbf{x}_n=\mathbf{x}+t_n\boldsymbol{\epsilon}$, with $\boldsymbol{\epsilon}\sim\mathcal{N}(0,\boldsymbol{I})$.

\section{Methods and Algorithms} \label{sec:methods}
\subsection{Threat Model}
Due to the increasing training cost, it is a common practice to use third-party models. Following \cite{chou2023backdoordiffusionmodels,chou2024villandiffusion}, there are two parties involved: (1) the user, who utilizes the off-the-self third-party model to perform a specific task, and (2) the attacker, who releases a compromised third-part model for malicious purposes. In a backdoor attack on consistency models, these models behave in a predetermined way when a specific trigger is activated and behave normally otherwise. Specifically, they generate clean image when sampling from Gaussian noise, but produce backdoor targets once a trigger is added to the noise. A trigger is a predefined image.

There are two types of metrics to measure the quality of the models: utility and specificity. Utility metrics, such as Fréchet Inception Distance (FID) \cite{FID} and Inception Score (IS) \cite{inception_score(IS)}, assess the quality of the generated images. Specificity metrics, like Mean Square Error (MSE), used solely by the attacker, evaluate how similar the generated backdoor images are to the backdoor target. The attacker will release the backdoor-compromised model if both the utility and specificity meet the requirements, while the user will accept the model if its utility meets their requirements.

In the backdoor attack scenario, we consider a white-box attack, where the attacker has full access to the training data, training schedule, hyperparameters and loss function. In contrast, the user has access only to the released model parameters and a subset of the \emph{clean} training data, allowing them to evaluate the model's performance.

\begin{algorithm}[t]
    \caption{Backdoor Consistency Model Training}
    \label{alg:training}
    \textbf{Input}: clean dataset $\mathcal{D}_c$, backdoor target dataset $\mathcal{D}_p$, pretrained diffusion/consistency model $\boldsymbol{\theta}_0$, log-normal distribution $p(t)$, mapping function $p(r|t,k)$, weighting function $\lambda(t)$, $\mathbf{R}$ as defined in Eq. (\ref{eq:R}), and poison rate $\rho$.
    
    \begin{algorithmic}[1]    
    
    \STATE $\boldsymbol{\theta}\leftarrow\boldsymbol{\theta}_0, k=0$
    \REPEAT
    \STATE Sample $\boldsymbol{\epsilon}\sim\mathcal{N}(0,\boldsymbol{I}), t\sim p(t), r\sim p(r|t,k)$
    \STATE Sample $u$ from a uniform distribution $\mathcal{U}[0,1]$,
    \IF{$u<\rho$}
        \STATE Sample $\mathbf{x}'_0\sim\mathcal{D}_p$
        \STATE Compute $\mathbf{x}_t=\mathbf{x}'_0+t\mathbf{R}+t\boldsymbol{\epsilon}$
        \STATE Compute $\mathbf{x}_r=\mathbf{x}'_0+r\mathbf{R}+r\boldsymbol{\epsilon}$
    \ELSE
        \STATE sample $\mathbf{x}_0\sim\mathcal{D}_c$
        \STATE Compute $\mathbf{x}_t=\mathbf{x}_0+t\boldsymbol{\epsilon}$
        \STATE Compute $\mathbf{x}_r=\mathbf{x}_0+r\boldsymbol{\epsilon}$
    \ENDIF
    \STATE $\mathcal{L}(\boldsymbol{\theta}, \boldsymbol{\theta^-})=\lambda(t)\cdot d(\boldsymbol{f_\theta}(\mathbf{x}_t),\boldsymbol{f_{\theta^-}}(\mathbf{x}_r))$
    \STATE $\boldsymbol{\theta}\leftarrow\boldsymbol{\theta}-\eta\nabla_{\boldsymbol{\theta}}\mathcal{L}(\boldsymbol{\theta},\boldsymbol{\theta^-})$
    \STATE $k=k+1$
    \UNTIL convergence
    
    \end{algorithmic}
\end{algorithm}

\subsection{Backdoor Consistency Models} \label{subsec: backdoor CM}
For training clean image consistency models, as discussed in Section \ref{sec:preliminary}, the training schedule is defined as
\begin{equation}
    \mathbf{x}_t = \mathbf{x} + t\epsilon, \epsilon\sim\mathcal{N}(0,\boldsymbol{I}),
\end{equation}
where $\mathbf{x}$ is the training image sampled from $p_{data}(\mathbf{x})$ and $\mathbf{x}_t$ is the sample at time $t$.

Correspondingly, we propose that the backdoor consistency model training schedule be defined as
\begin{equation} \label{eq:backdoor_sched}
    \mathbf{x}'_t = \mathbf{x}' + t\mathbf{R} + t\epsilon, \epsilon\sim\mathcal{N}(0,\boldsymbol{I}),
\end{equation}
 where $\mathbf{x}'$ is the backdoor target sampled from the target distribution $p_{target}(\mathbf{x}')$, $\mathbf{x}'_t$ is the sample at time $t$, and
 \begin{equation}\label{eq:R}
     \mathbf{R}=\mathbf{M}\odot \mathbf{g}+(1-\mathbf{M})\odot \mathbf{x},
 \end{equation}
 where $\mathbf{g}$ is the trigger, and $\mathbf{M}\in \{0,1\}$ is a binary mask.

We propose that the backdoor training loss function be defined as
\begin{align}
& \mathcal{L}'(\boldsymbol{\theta},\boldsymbol{\theta^-}) =\notag \\
& \mathbb{E}[\lambda(t_n)d(\boldsymbol{f_\theta}(\mathbf{x}'_{n+1}, t_{n+1}), \boldsymbol{f_{\theta^-}}(\mathbf{x}'_n,t_n))],
\end{align}
where $\mathbf{x}'_{n+1}=\mathbf{x}'+t_{n+1}\mathbf{R}+t_{n+1}\boldsymbol{\epsilon}$, $\mathbf{x}'_n=\mathbf{x}'+t_n\mathbf{R}+t_n\boldsymbol{\epsilon}$, and other symbols are defined in Eq. (\ref{eq:loss}).

During the consistency model training, the two adjacent times $t_{n+1}$ and $t_{n}$ are sampled as suggested in \cite{geng2024consistencymodelseasy}. Specifically, $t_{n+1}$ is sampled from a distribution $p(t)\sim\mathrm{LogNormal}(\mu,\sigma)$, while $t_n$ is sampled from a mapping function $p(r|t)$. Initially, $r=0$; As training progresses, $\frac{r}{t}\rightarrow1$. Readers are referred to \cite{geng2024consistencymodelseasy} for detailed parameter settings and the mapping function designs. 
The backdoor training algorithm is summarized in Algorithm \ref{alg:training}.

\begin{figure*}[h] 
     \centering
     \includegraphics[width=0.8\textwidth]{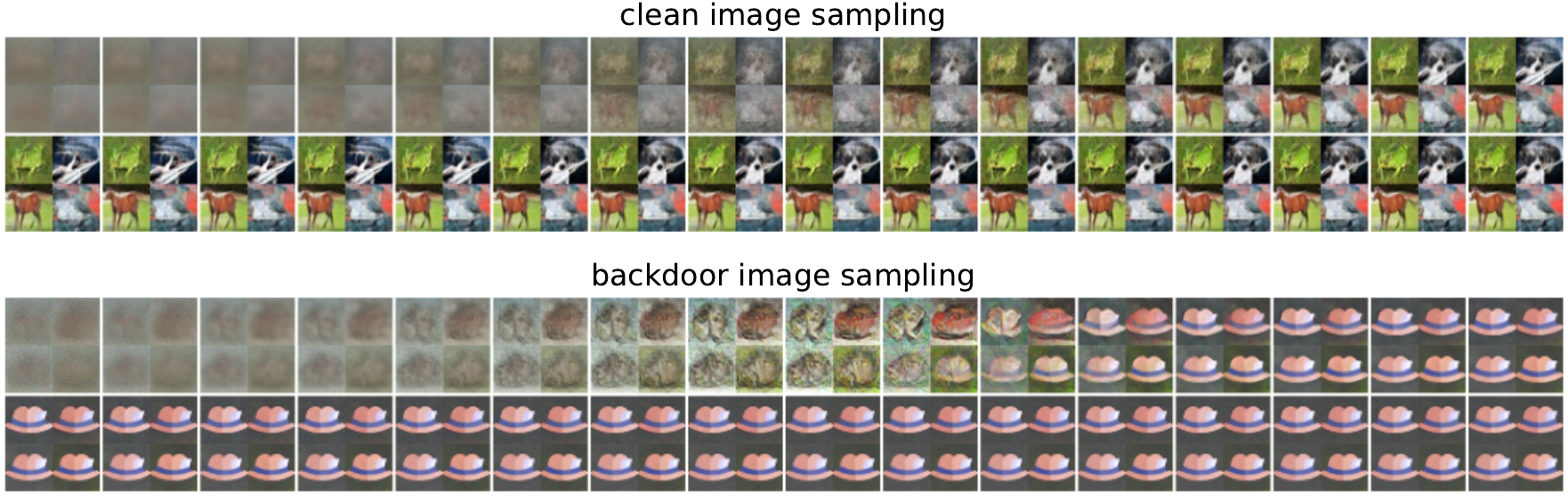} 
     \caption{Sampling of clean and backdoor images as training progresses (fine-tuning a pre-trained \emph{diffusion} model on the CIFAR-10 dataset).}
     \label{fig:progress-noise-hat-DM}
\end{figure*}
\begin{figure*}[h] 
     \centering
     \includegraphics[width=0.8\textwidth]{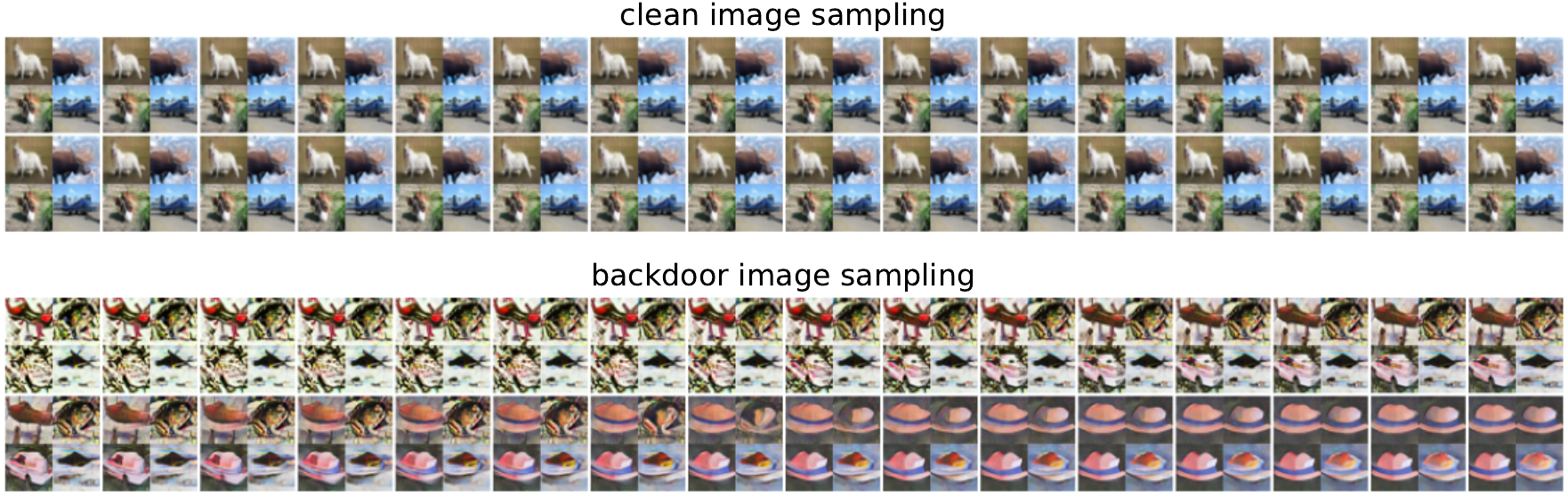} 
     \caption{Sampling of clean and backdoor images as training progresses (fine-tuning a pre-trained \emph{consistency} model on the CIFAR-10 dataset). Note that the clean image sampling exhibits no visual difference since the training starts from a well-trained clean \emph{consistency} model.}
     \label{fig:progress-noise-hat-cm}
\end{figure*} 

\begin{figure*}[h] 
     \centering
     \includegraphics[width=0.95\textwidth]{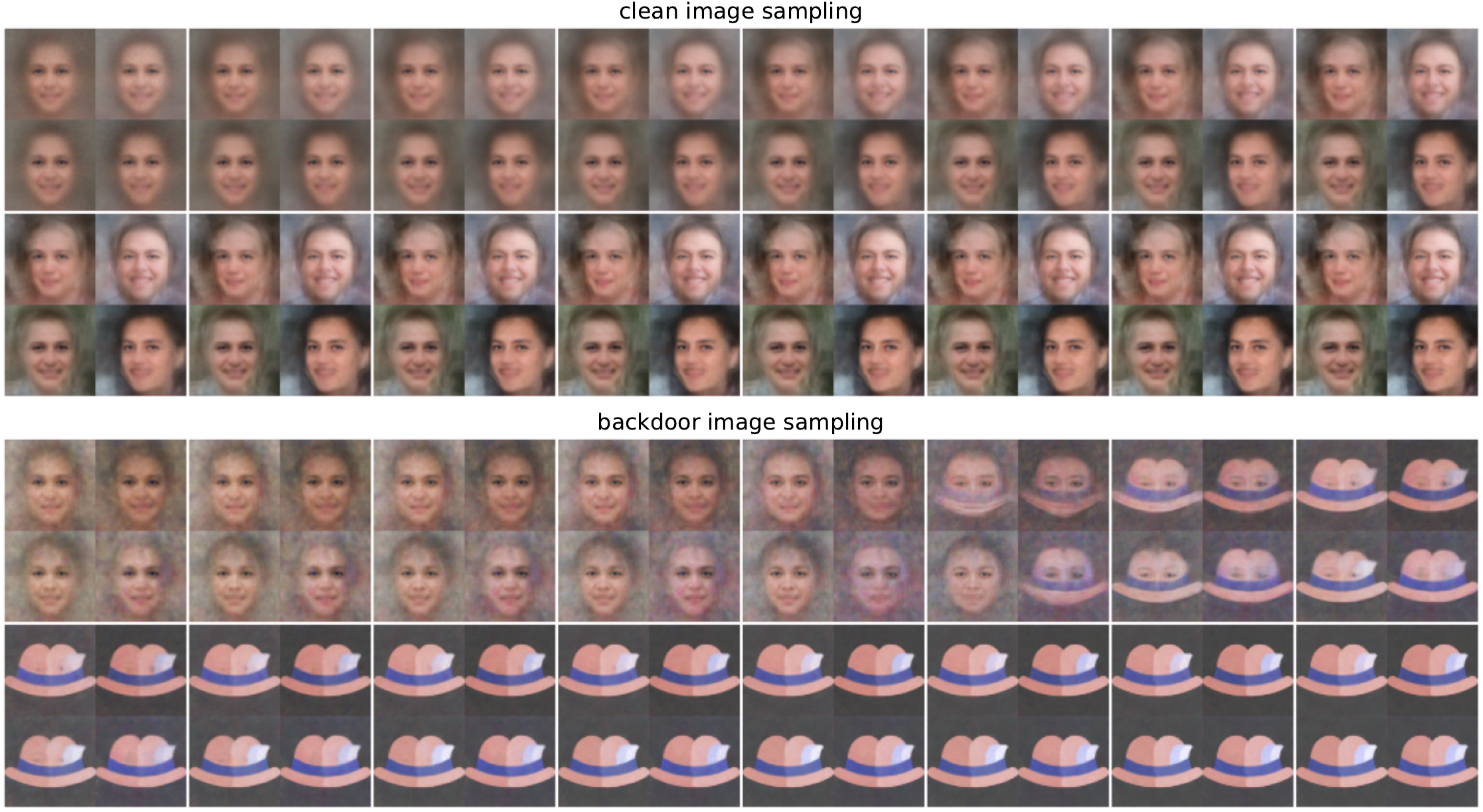} 
     \caption{Sampling of clean and backdoor images as training progresses (fine-tuning a pre-trained \emph{diffusion} model on the FFHQ dataset). Note that the clean images samples are initially blurred, but their quality progressively improves with further training, eventually reaching a level comparable to samples generated from a clean CM model (not shown here).}
     \label{fig:progress-noise-hat-DM-ffhq}
\end{figure*}
\begin{table}[]
\caption{Triggers and targets}
\centering
\begin{tabular}{ccc|cc}
\hline
\multicolumn{3}{c|}{Triggers}            & \multicolumn{2}{c}{Targets}    \\ \hline
\multicolumn{1}{c|}{Noise} & Box & Glasses & \multicolumn{1}{c|}{Hat} & Cat \\ 
\multicolumn{1}{c|}{\includegraphics[width=1cm]{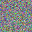}} & \includegraphics[width=1cm]{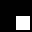} & \includegraphics[width=1cm]{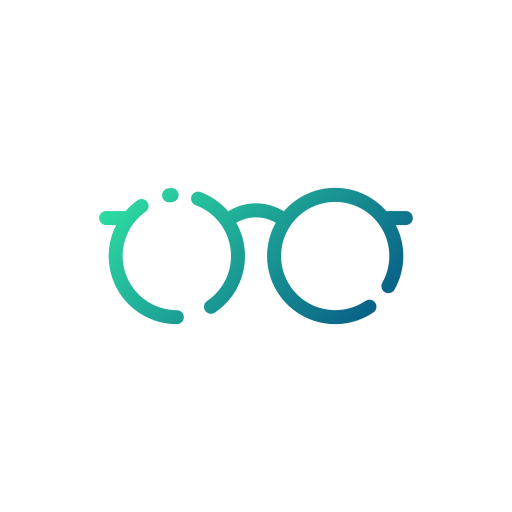} 
& \multicolumn{1}{c|}{\includegraphics[width=1cm]{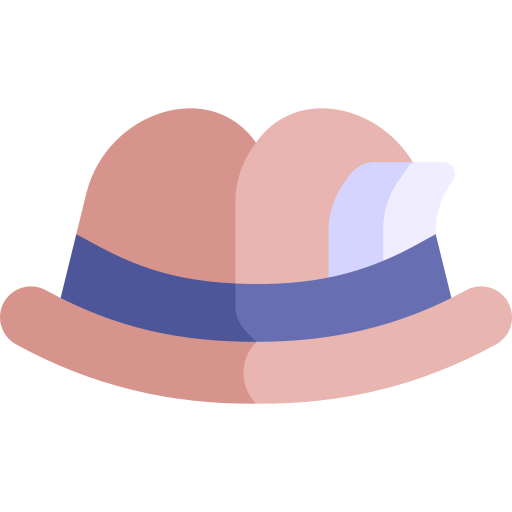}} & \includegraphics[width=1cm]{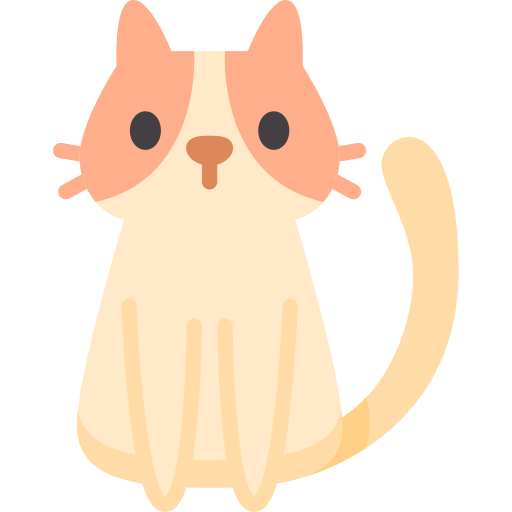}  \\ \hline
\end{tabular}
\label{tab:triggers_targets}
\end{table}




\section{Experiments} \label{sec:exp}
\subsection{Experimental setup}
We use the benchmark dataset CIFAR-10 $(32\times32)$ \cite{cifar10} and the FFHQ \cite{karras2019style_ffhq} dataset, which is converted to a resolution of $64\times64$. In the following text, CIFAR-10 is used as the default dataset, unless specified otherwise.

We use Box and Glasses as triggers, and Hat and Cat as targets, following \cite{chou2023backdoordiffusionmodels}. Additionally, we introduce noise as a trigger, which is less noticeable since consistency models sample from Gaussian noise when generating images. The noise trigger is obtained by sampling from a Gaussian distribution, and is fixed throughout the backdoor training process. The triggers and targets used in this paper are summarized in Table \ref{tab:triggers_targets}. By default, noise is used as the trigger and Hat as the target, unless specified otherwise.

We train the backdoor consistency model by \emph{fine-tuning a pre-trained base model}, which can be either a diffusion model or a consistency model~\cite{karras2022elucidating,geng2024consistencymodelseasy}. We use a pre-trained diffusion model by default. The poison rate, defined as the fraction of training time allocated for backdoor training, is set to 0.1 by default.

\subsection{Evaluation Metrics}
We use Fréchet Inception Distance (FID) \cite{FID} to measure the utility of the generated clean images. The FID is calculated between 50,000 generated clean images and all available training images,
which also totals 50,000 for the CIFAR-10 dataset. A lower FID indicates better quality of the generated images.

To assess the specificity of the generated backdoor images, we use Mean Square Error (MSE). MSE is computed as the mean pixel value difference between 64 generated backdoor images and the ground truth targets. A lower MSE indicates better backdoor image generation.



\subsection{Generating Samples as Training Progresses}
We visualize the generated clean images and backdoor targets as training progresses on the CIFAR-10 dataset, as shown in Figure \ref{fig:progress-noise-hat-DM}. The default noise trigger is used and the base model is a pre-trained \emph{diffusion} model. It is evident as training continues, the consistency model improves, gradually learning the data distribution.

We also visualize the generated clean images and backdoor targets as training progresses during the fine-tuning of a pre-trained \emph{consistency} model on the CIFAR-10 dataset, as illustrated in Figure \ref{fig:progress-noise-hat-cm}. The generated clean image exhibit no visual difference since \emph{the base model is a pre-trained consistency model}. It is observed that the backdoor training in this setting converges more slowly.

For the FFHQ dataset, we visualize the generated clean images and backdoor targets as training progresses during the fine-tuning of a pre-trained \emph{diffusion} model, as shown in Figure \ref{fig:progress-noise-hat-DM-ffhq}. Due to space constraints, the visualization of the training process for fine-tuning a pre-trained \emph{consistency} model is included in the Supplement.


\begin{figure}[h]
     \centering
     \includegraphics[width=0.45\textwidth]{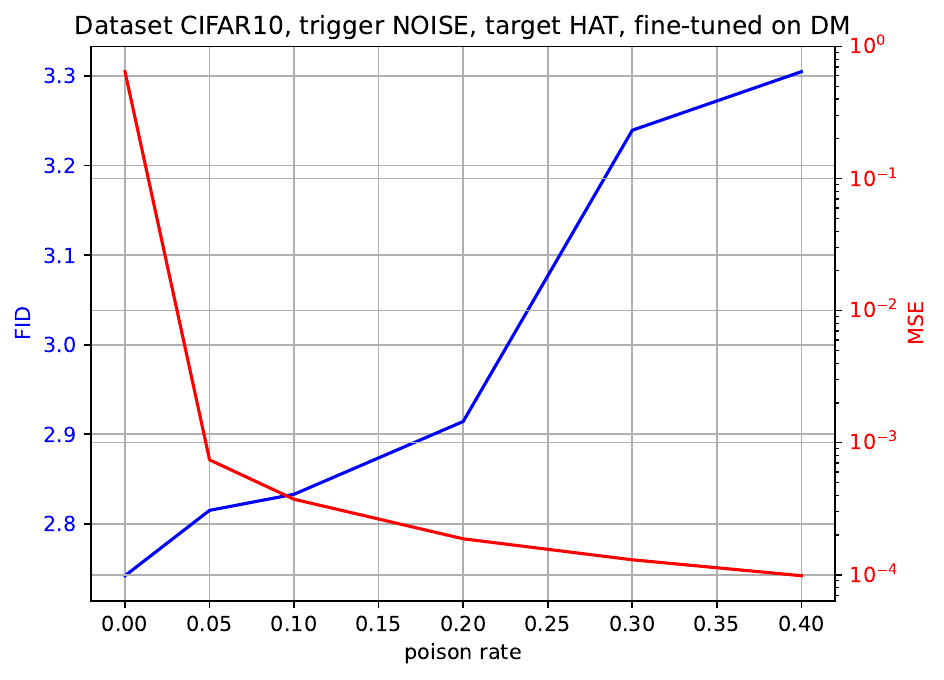} 
     \caption{FID and MSE values with respect to poison rate, fine-tuning on a pre-trained \emph{diffusion} model.}
     \label{fig:FID_MSE-noise-hat}
\end{figure} 
\begin{figure}[h]
     \centering
     \includegraphics[width=0.45\textwidth]{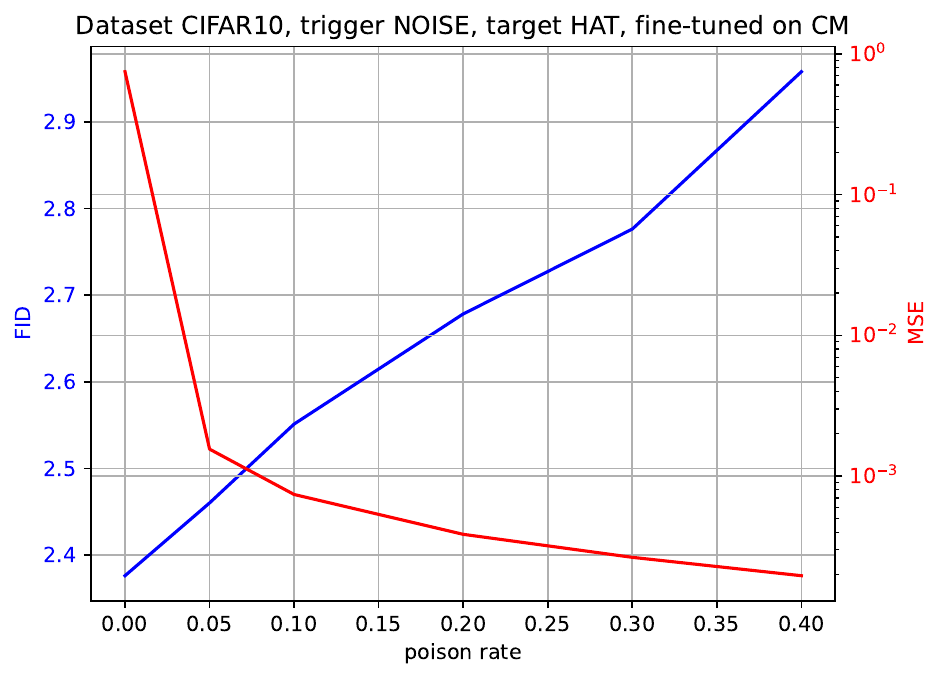} 
     \caption{FID and MSE values with respect to poison rate, fine-tuning on a pre-trained \emph{consistency} model.}
     \label{fig:FID_MSE-noise-hat-cm}
\end{figure} 

\subsection{Varying Poison Rate}
We investigate how the utility and specificity vary with different poison rates under various trigger and target settings. We combine FID and MSE into a single figure, as shown in Figure \ref{fig:FID_MSE-noise-hat}. The poison rate for clean images is set to 0. As the poison rate increases, the FID rises, indicating a gradual decline in the quality of the generated images. When the poison rate reaches 0.4, the FID increases by approximately 20\%, signaling a significant drop in quality. Therefore, we did not record the FID for higher poison rates. Conversely, as the poison rate increases, the MSE decreases, indicating a gradual improvement in the quality of the generated backdoor images. The data reveal that the FID and MSE values are well-balanced at poison rates ranging from 0.05 to 0.2 for this setting. Additionally it is evident that a $5\%$ poison rate can successfully backdoor a clean consistency model.

Figure \ref{fig:FID_MSE-noise-hat-cm} illustrates the FID and MSE values obtained by fine-tuning a pre-trained \emph{consistency} model, revealing a similar pattern. It is observed the balanced poison rate in this setting ranges from $0.05$ to $0.1$, where we achieve both low MSE and low FID, resulting in high utility and high specificity. Note that it can achieve a lower FID score under the same poison rate.

Additional experimental results with various trigger and target combinations can be found in the Supplement, which demonstrate a \emph{similar pattern}.

\section{Defense}\label{defense}
Several studies have proposed countermeasures against backdoor attacks on diffusion models \cite{an2024elijah,sui2024disdet,mo2024terd}. Due to the unique properties of consistency models and the stealthiness of the Gaussian noise trigger, most of them do not work in our setting. For instance, \cite{sui2024disdet} observes that the distribution of poisoned noise containing a trigger differs from that of clean noise. However, this assumption does not hold for our novel Gaussian noise trigger, which shares the same distribution as clean noise. The insight of \cite{an2024elijah} is that a backdoored diffusion model exhibits a distribution shift, which is dependent on the backdoor trigger. Specifically, if the input at time $t$ contains a trigger, the output of the compromised diffusion model will include an additional backdoor term at time $t-1$, where $t-1>0$ represents an intermediate step. In contrast, the output of consistency models is a clean images at time $t=0$, making the Eq. (4) in their paper inapplicable to backdoored consistency models. 

The defense proposed in \cite{mo2024terd} introduces an optimization method to approximate the hidden trigger from pure noise, based on the insight that a backdoored model make better predictions from Gaussian noise combined with the hidden trigger than from Gaussian noise with random nonzero noise. While this method is likely to succeed in approximating the hidden trigger in diffusion models by manipulating their noise predictions, it is not directly applicable to consistency models, as consistency models predict clean images. Additionally, our novel Gaussian trigger can bypass the element-wise distribution-based detection method described in Eq. (22) of their work. This is because the distribution of two Gaussian variables remains Gaussian. Although the variance of the combined Gaussian distributions doubles, this can be mitigated by scaling down the value of $R$ and $\epsilon$ in Eq. (\ref{eq:backdoor_sched}).

\section{Limitations} \label{sec:limit}
The proposed backdoor attack is a white-box attack, requiring the attacker to have access to the training code. The attack could be an insider within an organization or may release a compromised model directly, making the attack feasible in these scenarios.

\section{Conclusion} \label{sec:conclusion}
In this work, we present the first backdoor attack tailored specifically for consistency models. Our extensive experimental results clearly demonstrate the vulnerability of these models. Given the inherent strengths of consistency models, their potential for wide applications, and the stealthiness introduced by combining consistency models with a Gaussian noise trigger, our findings underscore the potential risks associated with applications built upon these frameworks.

\bibliographystyle{apalike}
\bibliography{references}


 \newpage
\onecolumn

\section*{How to Backdoor Consistency Models?: Supplementary Materials}

\section{Implementations} \label{supp:impl}
The implementation is based on the code from \cite{geng2024consistencymodelseasy}\footnote{https://github.com/locuslab/ect}. We create the backdoor triggers and targets based on the code from \cite{chou2024villandiffusion}\footnote{https://github.com/IBM/VillanDiffusion}.

The clean and backdoor images for the CIFAR10 dataset shown in Figure \ref{fig:progress-noise-hat-DM}, \ref{fig:progress-noise-hat-cm} are sampled at each tick, starting from tick 0 and ending at tick 31, with each tick iterating over $12,800$ images during model training. 

\begin{figure*}[h] 
     \centering
     \includegraphics[width=0.95\textwidth]{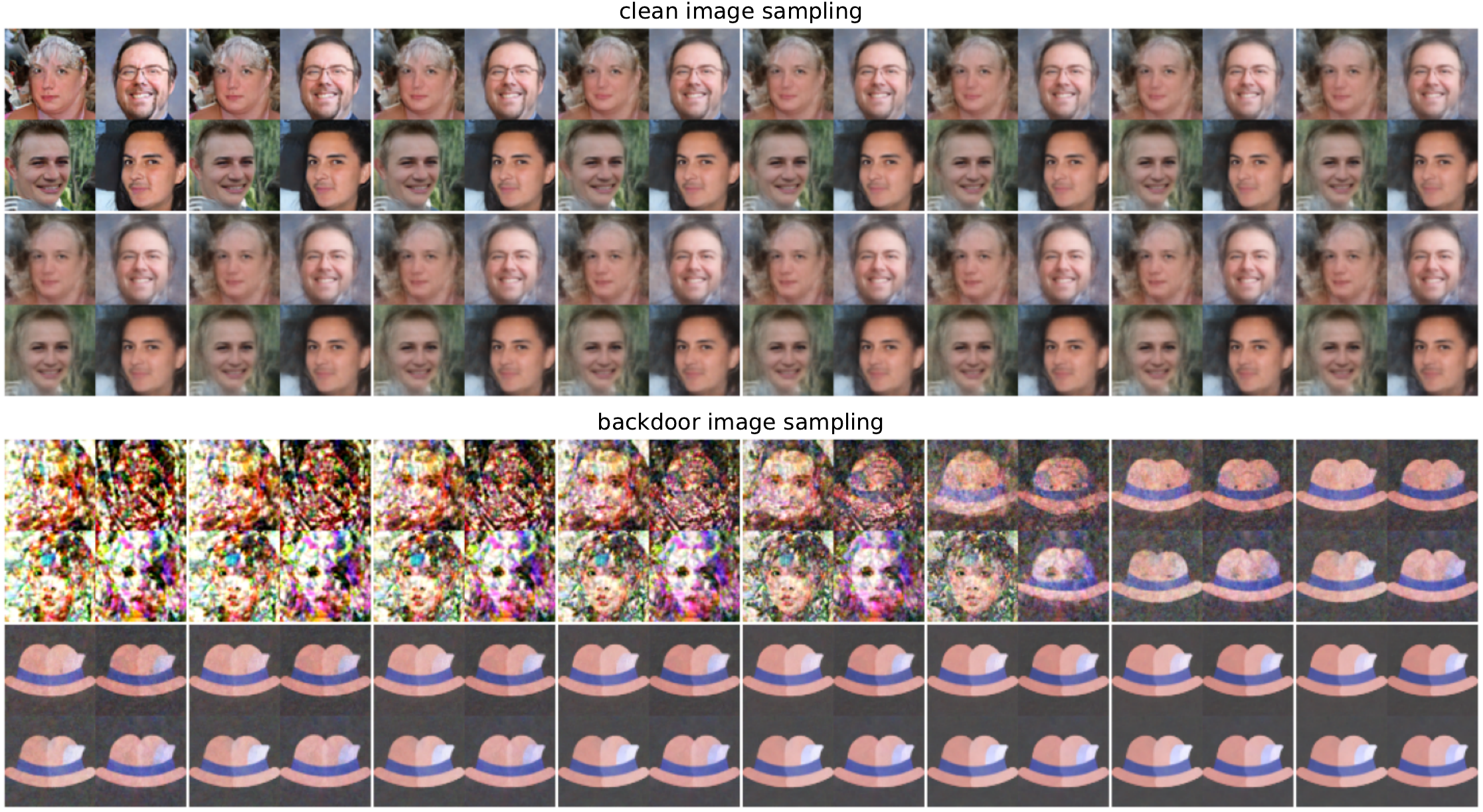} 
     \caption{Sampling of clean and backdoor images as training progresses (fine-tuning a pre-trained \emph{consistency} model on the FFHQ dataset). Note that the clean images samples exhibit quality degradation as training progresses, but their quality eventually returns to a level similar to that of samples generated from a clean CM model with further training (not shown here).}
     \label{fig:progress-noise-hat-cm-ffhq}
\end{figure*}

The clean and backdoor images for the FFHQ dataset shown in Figure \ref{fig:progress-noise-hat-DM-ffhq}, \ref{fig:progress-noise-hat-cm-ffhq} are sampled at every 10 ticks, starting from tick 0 and ending at tick 310, with each tick iterating over $12,800$ images during model training.

The FID and MSE values of all the models, fine-tuned on a diffusion model, are calculated after 240 ticks. The FID and MSE values of all the models, fine-tuned on a consistency model, are calculated after 250 ticks.

Due to the limitations in computing resources, the FID and MSE values are obtained by running the code once for each poison rate. These values are generally stable at a fixed poison rate.

The default setting for $M$ in Eq. \ref{eq:R} is set to all 1.

\section{Additional Experimental Results}

\subsection{Generating Samples as Training Progresses}
We visualize the backdoor training process for the FFHQ dataset in Figure \ref{fig:progress-noise-hat-cm-ffhq}. This model is fine-tuned on a pre-trained \emph{consistency} model.

\begin{figure}[ht]
    \centering
    \begin{subfigure}[b]{0.45\textwidth}
        \centering
        \includegraphics[width=\textwidth]{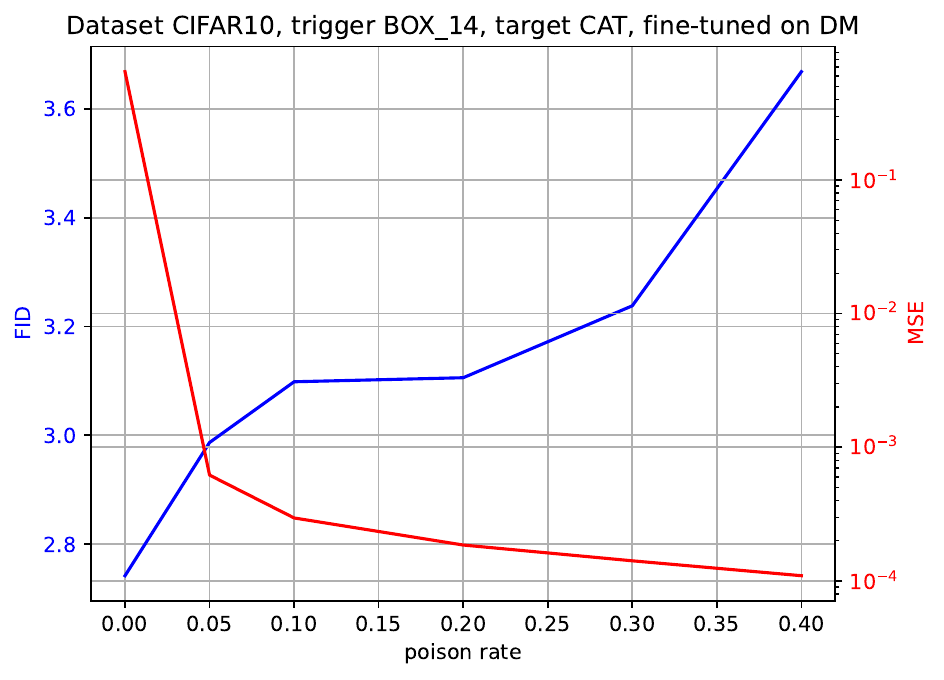} 
        \caption{}
        \label{fig:fig1}
    \end{subfigure}
    \hspace{0.05\textwidth} 
    \begin{subfigure}[b]{0.45\textwidth}
        \centering
        \includegraphics[width=\textwidth]{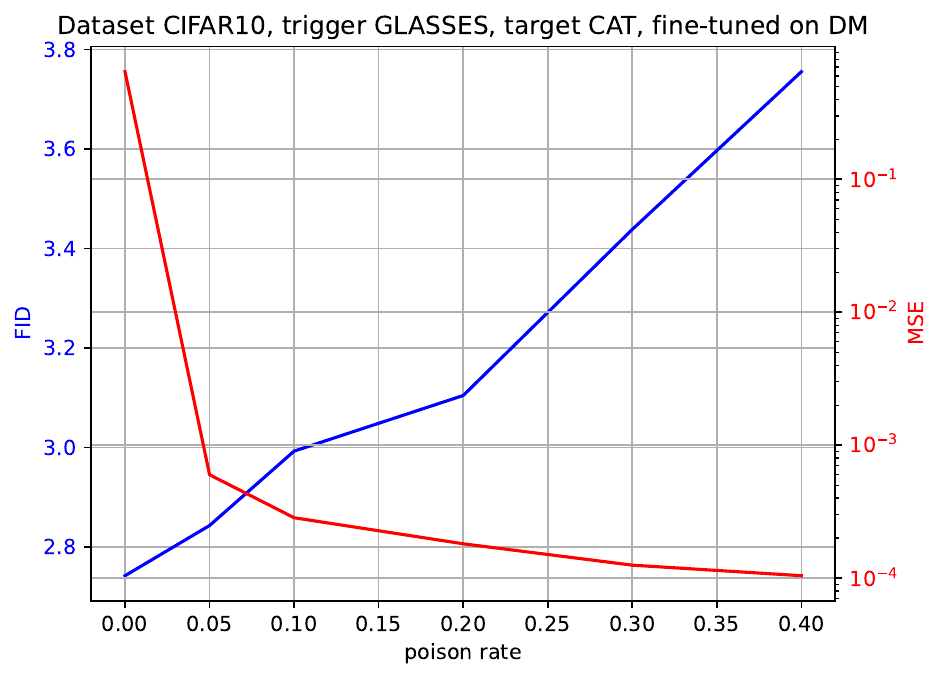} 
        \caption{}
        \label{fig:fig2}
    \end{subfigure}
    \caption{FID and MSE values with respect to poison rate, fine-tuning on a pre-trained diffusion model.}
    \label{fig:one column_DM}
\end{figure}
\begin{figure}[ht]
    \centering
    \begin{subfigure}[b]{0.45\textwidth}
        \centering
        \includegraphics[width=\textwidth]{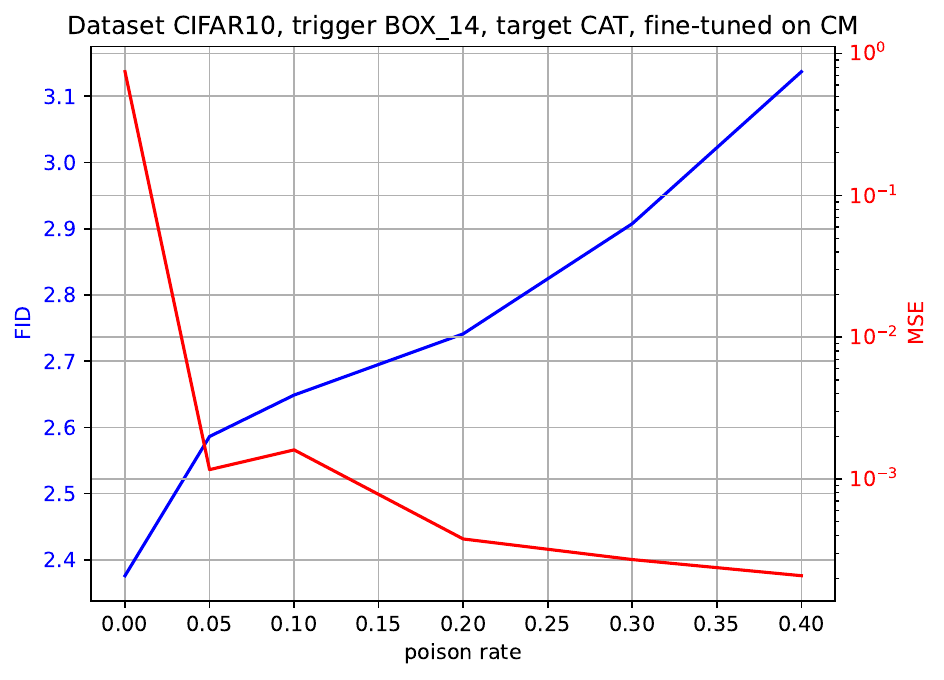} 
        \caption{}
        \label{fig:fig1}
    \end{subfigure}
    \hspace{0.05\textwidth} 
    \begin{subfigure}[b]{0.45\textwidth}
        \centering
        \includegraphics[width=\textwidth]{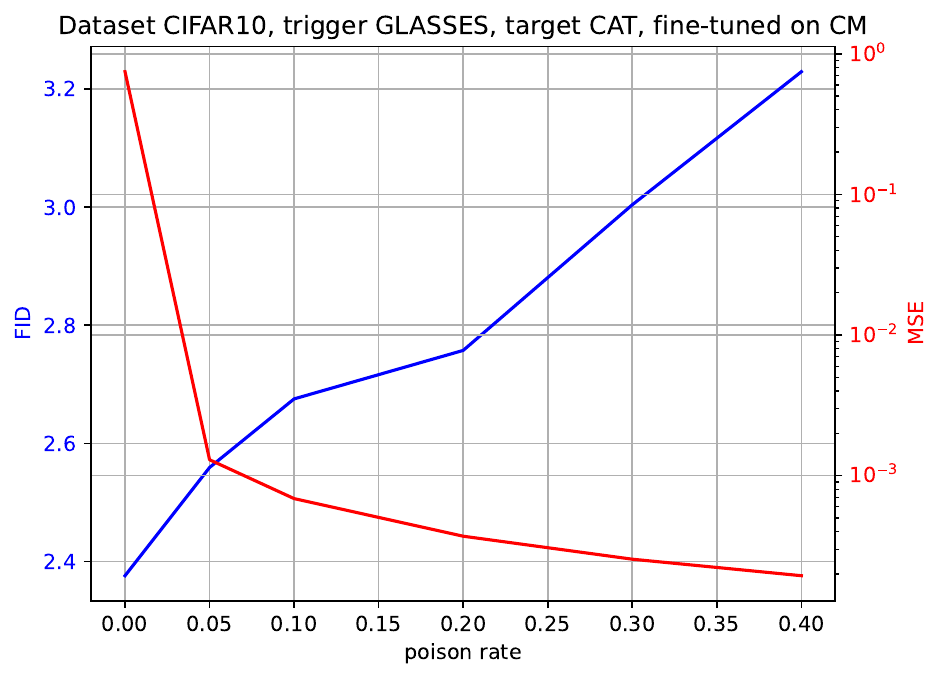} 
        \caption{}
        \label{fig:fig2}
    \end{subfigure}
    \caption{FID and MSE values with respect to poison rate, fine-tuning on a pre-trained consistency model.}
    \label{fig:one column_CM}
\end{figure}

\subsection{Varying Poison Rate} \label{supp:varying rate}

Due to limited computing resources, we report experimental results only for the CIFAR-10 dataset.

Figure \ref{fig:one column_DM} illustrates how the FID and MSE values vary across different poison rate, where the backdoor consistency models are trained from a pre-trained \emph{diffusion} model, with different triggers and targets.

Figure \ref{fig:one column_CM} illustrates how the FID and MSE values vary across different poison rate, where the backdoor consistency models are trained from a pre-trained \emph{consistency} model, with different triggers and targets.

These figures exhibit a pattern consistent with that described in the main text.

\end{document}